# Balancing Usability and Compliance in AI Smart Devices: A Privacy-by-Design Audit of Google Home, Alexa, and Siri

Trevor De Clark
*Computer Science Department*
*Vancouver Island University*
Nanaimo, Canada
trevor.declark@viu.ca

Yulia Bobkova
*Computer Science Department*
*Vancouver Island University*
Nanaimo, Canada
yulia.bobkova@viu.ca

Ajay Kumar Shrestha
*Computer Science Department*
*Vancouver Island University*
Nanaimo, Canada
ajay.shrestha@viu.ca

***Abstract*— This paper investigates the privacy and usability of AI-enabled smart devices commonly used by youth, focusing on Google Home Mini, Amazon Alexa, and Apple Siri. While these devices provide convenience and efficiency, they also raise privacy and transparency concerns due to their always-listening design and complex data management processes. The study proposes and applies a combined framework of Heuristic Evaluation, Personal Information Protection and Electronic Documents Act (PIPEDA) Compliance Assessment, and Youth-Centered Usability Testing to assess whether these devices align with Privacy-by-Design principles and support meaningful user control. Results show that Google Home achieved the highest usability score, while Siri scored highest in regulatory compliance, indicating a trade-off between user convenience and privacy protection. Alexa demonstrated clearer task navigation but weaker transparency in data retention. Findings suggest that although youth may feel capable of managing their data, their privacy self-efficacy remains limited by technical design, complex settings, and unclear data policies. The paper concludes that enhancing transparency, embedding privacy guidance during onboarding, and improving policy alignment are critical steps toward ensuring that smart devices are both usable and compliant with privacy standards that protect young users.**

## I. Introduction

The rapid integration of AI smart-devices into our everyday lives and homes has ushered in a new era of ever-present computing, with AI-powered smart devices like Google Home, Amazon Alexa, and Apple Siri becoming deeply embedded in the domestic and personal lives of millions, including youth who are among the earliest and most avid adopters of new technologies [1], [2]. These voice-activated assistants offer unparalleled conveniences, from small tasks like setting a timer to controlling entire smart home ecosystems, fundamentally changing how users interact with technology [3]. However, this convenience comes at a cost, posing unique privacy and transparency risks for youth who may be less equipped to understand the long-term implications of data exposure [4]. The always-on, data-intensive nature of voice-activated devices raises significant concerns related to covert data capture, unauthorized surveillance, and opaque data-sharing practices [5], [6].

In response to these systematic risks, Privacy-by-Design (PbD) has emerged as a critical guiding principle, advocating for privacy to be embedded into software architecture and business practices by default, and not as an afterthought [7]. This principle is increasingly reflected in global regulatory frameworks such as the European Union's General Data Protection Regulation (GDPR) and has been proposed as a core amendment to Canada's Personal Information Protection and Electronic Documents Act (PIPEDA) [8], [9]. A significant increase in technical research has suggested PbD-aligned solutions such as federated learning and on-device processing to minimize data exposure [10], [11]. At the same time, studies have also highlighted the unique vulnerabilities of young users, who often struggle with convoluted consent models and face a "privacy paradox" where expressed concerns do not align with their digital behaviors [12], [13], [14]. The gap between conceptual protection and practical user experience (UX) highlights the importance of implementing PbD as an operational rather than purely policy-level framework [15].

Despite this growing body of literature, a significant research gap persists. Existing studies have synthesized privacy risks, policy shortcomings, and possible technical advancements at a conceptual level. However, few studies have taken a practical approach to evaluating device-level audits to assess whether these PbD principles and regulatory requirements are implemented in the consumer smart devices people use daily. There remains a lack of integrated evaluation that examines the usability of privacy controls, transparency of data collection practices, and compliance with legal standards such as PIPEDA. This paper addresses this gap by conducting a comparative privacy audit of three major AI-enabled smart devices: Google Home, Amazon Alexa, and Apple Siri, to assess usability, transparency, and compliance with Canadian privacy standards.

The study is guided by the three research questions: How usable are device privacy controls for common actions such as disabling voice history, deleting stored commands, and verifying changes? To what extent do these devices align with PIPEDA's principles and the Office of the Privacy Commissioner's guidance on meaningful consent? Where do youth users encounter friction or uncertainty when attempting to manage their privacy on these devices?

The remainder of the paper is structured as follows: Section II provides the background and outlines the context of the study.

Section III details the methods used in the audit. Results are presented in Section IV, with the discussion and limitations following in Section V. Finally, Section VI concludes the paper.

## II. Background and Related Work

The rapid adoption of AI-enabled smart devices has introduced a new array of privacy challenges, particularly for young users. Voice-activated devices like those examined in this audit employ an "always-listening" architecture that creates inherent surveillance risks, including unintentional activation and covert data capture [16], [17]. This architecture creates an "oxymoron" of surveillance within the supposed privacy of the home [18]. These devices frequently collect sensitive behavioural data transmitted to cloud servers, creating vulnerabilities to cybersecurity breaches [6], [19]. Young digital citizens face compound risks arising from development factors and the well-documented privacy paradox; their stated concerns about privacy often fail to translate into proactive protective behaviours [13], [20]. Studies consistently suggest that young users struggle with complex consent mechanisms and lengthy privacy policies, leaving them at risk for data exploitation [21], [22], [23], [24]. This is exacerbated by a general lack of understanding of data ownership and the commercial infrastructures behind digital services [25].

### A. *Regulatory Frameworks: PIPEDA and the Challenge of Meaningful Consent*

Canada's privacy landscape is governed by the PIPEDA, which establishes ten information-handling principles: accountability, identifying purposes, consent, limiting collection, limiting use, accuracy, safeguards, openness, individual access, and challenging compliance [26]. These principles mandate that the organization be accountable for personal data, obtain meaningful consent for its collection, and use it only for specific, legitimate purposes. They must limit data collection, ensure accuracy, and safeguard information with appropriate safety measures. The principles outline users' right to access their data and challenge an organization's compliance, while requiring the organization to be open and transparent about their privacy policies. The Office of the Privacy Commissioner's 2018 Guidelines for Obtaining Meaningful Consent further emphasize plain-language communication and user-friendly controls, particularly important for youth users. However, PIPEDA has been criticized for its lack of explicit, youth-specific safeguards, unlike stricter frameworks like the EU's GDPR, which embeds stronger defaults for minors [8], [27]. This regulatory gap highlights the importance of auditing whether devices themselves bridge this gap.

### B. *PbD and Technical Safeguards*

PbD has emerged as a critical framework for embedding privacy protections directly into the software [7]. The principle advocates for a proactive approach where privacy is a core feature of the system's design, ensuring data protection throughout the entire data lifecycle. Technical implications of PbD principles have become a significant focus of research, proposing various architectures to minimize data exposure. Federated Learning (FL) and decentralized architectures are an approach that enables AI models to be trained across multiple decentralized devices (e.g., user smartphones) using local data with only model updates, meaning no raw data is being sent to a centralized server for aggregation. This prevents the central collection of sensitive user data [28], [29]. Recent advancements have explored using blockchain-enhanced FL to further ensure data integrity and transparency in these distributed systems [30].

Edge processing has also emerged as a promising approach to reduce the risk of interception during the transmission of data, as it processes user data locally or on nearby edge servers rather than transmitting everything to the cloud. Techniques such as privacy-preserving deep learning on the edge have been developed to maintain functionality while keeping data local [11], [31]. Advanced cryptographic techniques are fundamental to PbD. Research has focused on lightweight encryption algorithms suitable for resource-constrained IoT devices [32], [33], as well as more sophisticated methods like homomorphic encryption, which allows computations to be performed on encrypted data without decrypting it first [34]. Furthermore, privacy-preserving authentication schemes, such as enhanced certificateless authentication, have been proposed to secure user access without compromising identity [35].

While these technical solutions show significant theoretical promise in academic research, their widespread implementation in current commercial consumer devices, particularly those used by youth and families, remains a significant challenge [36]. Moreover, even where such mechanisms may be partially adopted in proprietary platforms, they are typically opaque to end users: youth and their caregivers interact primarily with surface-level settings, permissions, and disclosures rather than underlying cryptographic or architectural choices. The gap between the proposed PbD mechanisms and their real-world deployment highlights the need for empirical audits to assess the current state of privacy protection in these devices. In this study, we therefore focus on how privacy-by-design principles manifest in observable user interfaces, configuration flows, and policy communications of mainstream smart assistants, rather than attempting to reverse engineer or verify the internal deployment of specific technical safeguards.

### C. *Evaluation Methods and Research Gap*

This study employs a Heuristic Evaluation, a usability engineering method adapted to assess privacy UX [37]. This approach enables expert review of privacy controls against established usability principles. Previous privacy assessment initiatives, such as Mozilla's Privacy Not Included guide, have provided valuable consumer-focused evaluations of smart devices. However, these efforts typically focus either on technical security analysis [6], [38], [39], [40], or policy and documentation review [16], [41], [42], [43], creating a fragmented understanding of actual privacy protections.

Consequently, a significant gap exists in the current literature. While current research addresses policy frameworks, technical vulnerabilities, or user perceptions on their own, few studies conduct integrated audits that simultaneously evaluate regulatory compliance, usability testing, and youth-centered design. This study addresses this gap by combining heuristic evaluation of privacy UX with PIPEDA compliance assessment and youth-focused task analysis, providing a comprehensive evaluation of whether privacy protections in everyday smart devices are both effective and accessible to vulnerable populations like young digital citizens.

## III. METHODOLOGY

This section outlines our device selection, evaluation framework, participant notes, and methods used for scoring and analysis.

### A. Device Selection

The devices were selected based on two key criteria: popularity among our targeted demographic of Canadian youths (ages 16-24) and platform diversity. Selecting devices from different platforms was essential to ensure accuracy and mitigate platform-based bias.

The devices used in this audit are the Google Home Mini (1st generation; Fuchsia OS 3.76 with Google Home companion app versions 3.39 and 3.4), Amazon Echo Dot (3rd generation; Fire OS 11040824196 with Alexa companion app versions 2.2.678637, 2.2.680179, 2.2.681009, and 2.2.683250), and Apple Siri on MacBook Air (M4 2025 running macOS Tahoe 26.0). All evaluations were conducted during September-October 2025, and devices were configured using region and language settings appropriate for Canadian users (e.g., Canadian service regions and English-language interfaces). Each device underwent a fresh setup using newly created accounts with no prior association to the platform, ensuring that default configurations and first-time privacy experiences were evaluated accurately.

### B. Evaluation Framework

To evaluate both usability and regulatory compliance, the study employed a structured three-part framework comprising a Heuristic Evaluation, a PIPEDA Compliance Checklist, and a series of UX tests. These complementary methods were selected to combine established usability testing approaches with regulatory compliance frameworks, ensuring methodological rigor and practical relevance.

The Heuristic Evaluation involved usability experts assessing an interface against a set of established design principles to identify potential usability problems. The evaluation is grounded in Nielsen and Molich's classic framework, widely adopted in human-computer interaction for identifying usability problems [37]. By tailoring these heuristics to privacy controls, the audit extends a validated technique to a contemporary privacy context, consistent with prior research on privacy-oriented design [44], [45]. This approach was used to assess the privacy settings screens of the three devices, with evaluators recording both score and qualitative notes for each of the *seven criteria* (Table I).

The PIPEDA Compliance Checklist assessed adherence to the ten *Fair Information Principles* (Table II) that form the basis of Canada's PIPEDA and was also informed by the Office of the Privacy Commissioner of Canada's *Guidelines for Obtaining Meaningful Consent (2018)* [26], [46]. This framework establishes recognized principles for transparency, consent, and control, ensuring the assessment criteria align with regulatory expectations. Each principle was scored based on compliance, with evidence (e.g., screenshots, menu paths, excerpts from policy text) used to justify scores.

Finally, the UX testing component evaluated how easily participants could locate, interpret, and modify privacy settings through *task-based scenarios* (Table III). To minimize prior-experience bias, participants were instructed to treat each device as if they were encountering its privacy settings for the first time. To mitigate order effects and learning bias, device presentation was counterbalanced using a Latin square, and task order within each device session was randomized per participant. Participants were instructed to verbalize uncertainties during tasks, which were recorded in observation notes but not scored. Average step counts are reported in the Results, and given the small sample size, they are interpreted qualitatively rather than as statistically robust measures.

This UX test component follows the well-established principle that small samples are sufficient to uncover the majority of usability problems. As articulated by Nielsen [47], testing with three to five participants typically reveals the most critical interface problems. In this study, youth participants provided illustrative, practice-based evidence of barriers to privacy management, complementing the expert-driven heuristic assessment. The small sample size is therefore appropriate given the exploratory, audit-oriented purpose of the evaluation rather than statistical generalization. After each task, participants rated how difficult they found the task and provided qualitative feedback on points of confusion; task completion time was measured separately using a stopwatch.

Data collection involved any relevant screen captures, observation notes, and usability scores. By triangulating expert heuristics, statutory compliance and participant usability testing, the PbD audit employs a mixed-method approach that is both validated and contextually adapted. The methodology is therefore an evidence-based integration of established practices, tailored to the realities of youth privacy in AI-enabled smart devices.

### C. Participant Note

The study received ethics approval from the Vancouver Island University Research Ethics Board (VIU-REB). The approval reference number #103597 was given for behavioral application/amendment forms, consent forms, interview and focus group scripts, and questionnaires.

To preserve anonymity, each participant was assigned a randomized identifier (DA-1 to DA-4). Participants were given information about their task to perform in advance and had the right to withdraw at any time. The audit involved four participants drawn from the target demographic of 16-24 years; all participants were between 18 and 24 years old, and no minors were included. This approach aligns with usability research, which finds that small sample sizes of 3-5 users are sufficient to uncover most interface and usability issues [47].

The purpose of this exercise was not to generate statistically generalizable findings but rather to provide illustrative, practice-based insights into the accessibility and transparency of privacy controls in smart devices. UX testing was conducted with three of the these four participants, while the remaining audit components, Heuristic Evaluation and PIPEDA Compliance Checklist, were conducted by a team of three evaluators with privacy-UX expertise. The evaluators independently reviewed each device's settings, account dashboards, and documentation. After independent scoring, the team conducted a consensus meeting to reconcile differences. Discrepancies greater than one point were discussed until agreement was reached; if disagreement persisted, the evaluators revisited the interface or documentation together and applied a shared rationale for the final score. This procedure follows established practice in heuristic and policy-compliance audits, which rely on structured expert review and explicit consensus protocols. The consensus-based scoring process helped ensure consistent application of evaluation principles across devices and provided a transparent rationale for the reported results.

### *D. Scoring and Analysis*

Each methodological component produced both quantitative and qualitative data to enable comparison across devices. A standardized 0-2 scale was used per criterion, where 0 = not met, 1 = partially met, and 2 = fully met. For Heuristic Evaluation, the total possible score was 14 points, representing seven heuristics evaluated. For the PIPEDA Compliance Checklist, the maximum possible score was 20 points, corresponding to the ten fair information-handling principles assessed. To ensure comparability, each device's heuristic and PIPEDA totals were normalized to a 0-1 range before computing the composite score (i.e., heuristic score ÷ 14; compliance score ÷ 20). The composite score is reported as the unweighted average of the two normalized values. Given the coarse 0-2 scale, several criteria approached the upper bound of the scoring range; the implications of this limited spread for cross-device comparison are discussed in Section IV. UX task data were not included in the composite score; they were incorporated qualitatively to contextualize the findings, highlight areas of friction, and explain discrepancies between usability and compliance metrics.

## IV. Results And Analysis

This section presents the findings structured around the three guiding questions: RQ1 (usability of privacy controls), RQ2 (alignment with PIPEDA principles), and RQ3 (youth experiences of friction and uncertainty when managing privacy on these devices). The PbD audit yielded quantitative scores and qualitative insights from the heuristic evaluations, PIPEDA compliance assessments, and youth UX testing.

The individual heuristics were applied as follows: Discoverability measured how easily the privacy settings could be located from the main menu; Comprehensibility assessed the clarity of the language used in the interface; Control evaluated the extent to which users could disable data collection for non-essential features; Granularity examined whether users could adjust permissions individually; Feedback considered whether the system confirmed the application of privacy changes (User Interface confirmation plus timely, consistent back-end enforcement and cross-context consistency); Reversibility assessed the ease with which privacy changes could be undone; and Consistency evaluated whether the settings were uniform across both the device and its companion application.

Addressing RQ1, the heuristic evaluation showed that the usability of privacy controls was comparable across devices. However, Google received a slightly higher score, primarily due to its superior feedback mechanisms compared to the other two devices. Table I summarizes the heuristic evaluation outcomes across the seven criteria for all three devices.

Discoverability is defined here as the visibility of a top-level access point to privacy settings from the main menu, without assessing the depth or navigational complexity of individual controls (e.g., "Disable voice history" path). Feedback, as applied in the heuristic evaluation, encompasses both the system's confirmation of a privacy change and the consistency with which that change is enforced and reflected across contexts. This definition extends beyond procedural verification measures, which are detailed in subsequent user testing sections. All three systems demonstrated strong discoverability and reversibility. Feedback mechanisms were weaker for Alexa and Siri, where changes experienced a delay or lacked visible confirmation. While each device provided a clear entry point to privacy settings, some critical controls, such as Siri's "Improve Siri & Dictation" toggle, located within the Privacy & Security settings under Analytics & Improvements, were placed outside the assistant's main configuration menu, potentially impacting user expectations and navigation.

In response to RQ2, the PIPEDA compliance audit examined ten information-handling principles: accountability, identifying purposes, consent, limiting collection, limiting use and disclosure, accuracy, safeguards, openness, individual access, and challenging compliance. Each criterion was scored based on the presence, accessibility, and clarity of the associated privacy mechanisms.

Accountability was assessed by determining whether the organization provided direct contact information for a privacy officer or equivalent team on its website; if not, the presence of a form to contact the privacy team was considered. Identifying Purposes evaluated whether the company included a statement detailing what data was collected and for what purpose. Consent examined whether users were required to opt in, rather than being automatically enrolled and needing to opt out. Limiting Collection considered whether the data collected was strictly necessary for the product's functionality. Limiting Use/Disclosure assessed whether users could restrict the sharing of their data with third parties. Accuracy evaluated whether users could review, edit, and delete their data. Safeguards examined whether encryption and related security measures were clearly described for handling user data. Openness considered whether the privacy policy was easy to locate and understand. Individual Access assessed whether users could download and delete their data. Finally, Challenging Compliance evaluated whether a clear process was available for filing complaints.

TABLE I. Heuristic Evaluation Results (3 devices * 7 criteria)

| Criterion | Heuristic Score by Device | | |
|---|---|---|---|
| | ***Google Home*** | ***Alexa*** | ***Siri*** |
| 1. Discoverability | 2 | 2 | 2 |
| 2. Comprehensibility | 2 | 2 | 2 |
| 3. Control | 2 | 2 | 2 |
| 4. Granularity | 2 | 2 | 2 |
| 5. Feedback | 2 | 1 | 1 |
| 6. Reversibility | 2 | 2 | 2 |
| 7. Consistency | 2 | 2 | 2 |
| Total | 14 | 13 | 13 |

[a] Heuristic Evaluation Results for Google Home, Amazon Alexa, and Apple Siri (Maximum Score = 14)

As shown in Table II, Siri achieved the highest total compliance score, reflecting stronger safeguards and greater transparency in consent and accuracy mechanisms. As a summary measure of overall performance, the normalized composite score for each device was computed as the average of its heuristic and PIPEDA subscores; for Google Home, this was calculated as (14/14+16/20)/2=0.90. The resulting composite scores were 0.90 for Google Home, 0.84 for Alexa, and 0.91 for Siri, capturing the balance of heuristic usability and PIPEDA compliance across devices. The heuristic evaluation exhibited a mild ceiling effect, as most heuristic items received scores of 1 or 2, consistent with mature, mainstream interfaces. By contrast, the PIPEDA checklist showed greater variability across principles such as Consent, Limiting Use and Disclosure, and Accuracy. As a result, the composite scores remained sensitive to meaningful differences in regulatory alignment even where usability scores were relatively high for all three platforms.

To address RQ3, the UX test evaluated the ease of use of the privacy settings interface. Each participant performed four standardized tasks on all devices: disabling voice recording history, finding data-sharing information, deleting stored commands, and verifying that a privacy change had taken effect. All participants reported basic familiarity with at least one of the three ecosystems (Google, Amazon, or Apple). A privacy modification was considered successfully verified when the interface confirmed that the change had been registered and would take effect. Participants recorded their perceived ease of completing each task. Ease-of-use was rated on a 1-5 scale, where 1 indicated very difficult, and 5 indicated very easy.

The number of steps was quantified by counting the total number of discrete button presses required to execute the specified action. It is important to note that, in certain smart device applications, a given function may be accessible or modifiable through multiple interface pathways. For example, a setting may be adjustable both via the companion application and through the user account configuration menu, or it may be reachable through submenu links. To account for variability in step count measurements arising from interface discrepancies, the number of steps required to complete each task was estimated by calculating the statistical mean across all individual participant step counts. This averaging procedure was employed to approximate the number of steps a typical user would likely perform when executing the task.

TABLE II. PIPEDA COMPLIANCE RESULTS (3 DEVICES * 10 PRINCIPLES)

| PIPEDA Principle | Smart Voice Assistants/Devices | | |
|---|---|---|---|
| | ***Google Home*** | ***Alexa*** | ***Siri*** |
| 1. Accountability | 1 | 1 | 1 |
| 2. Identifying Purposes | 2 | 2 | 2 |
| 3. Consent | 1 | 1 | 2 |
| 4. Limiting Collection | 2 | 1 | 2 |
| 5. Limiting Use/Disclosure | 2 | 1 | 2 |
| 6. Accuracy | 1 | 1 | 2 |
| 7. Safeguards | 2 | 2 | 2 |
| 8. Openness | 1 | 2 | 2 |
| 9. Individual Access | 2 | 2 | 1 |
| 10. Challenging Compliance | 2 | 2 | 2 |
| Total | 16 | 15 | 18 |

b. PIPEDA Compliance Checklist Scores (Maximum Score = 20)

Task completion time was measured using an external electronic device equipped with a stopwatch function. The timer was initiated when the participant began searching for the relevant setting and stopped upon successful modification of the target parameter. The timing error is estimated to be within $\pm 0.5$ seconds, which is negligible relative to overall task durations.

Among all tasks, disabling the voice recording history was consistently the most challenging across devices. Siri, in particular, posed significant difficulty for most participants, primarily because its settings were nested outside the Siri-specific configuration menu. Table III presents the average ease-of-use ratings, number of steps, and time taken for each task performed on each device.

## A. *Highlighted trends*

Google Home received the highest usability score among the three smart devices (14/14), as illustrated in Fig. 1. This was primarily due to its superior real-time feedback mechanism: a confirmation pop-up appears whenever a setting is changed, verifying that the change has been successfully applied. Aside from this feature, the settings interfaces across the devices were largely identical.

UX testing with the Google Home Mini revealed that certain voice assistant settings, such as disabling voice recording history, were managed through the user's Google account rather than directly within the Google Home device settings, potentially complicating access paths for inexperienced users. Apple's Siri exhibited similar usability challenges, particularly due to the distribution of configuration options across multiple settings interfaces. Participants in the UX test reported confusion regarding whether a desired setting resided within the general device settings or within Siri-specific settings, which consequently increased task completion time.

Siri outperformed the other two smart devices in the PIPEDA compliance evaluation, scoring highest overall (18/20). It demonstrated superior handling of consent by not automatically opting users into optional services. Siri also scored higher in Accuracy, supporting inline correction of transcribed queries (e.g., "Tap to Edit"), whereas Alexa and Google Home only allow users to view or delete stored interactions, with no option to edit the underlying transcripts or audio. A detailed comparison of Siri's compliance performance relative to Alexa and Google is presented in Fig. 2.

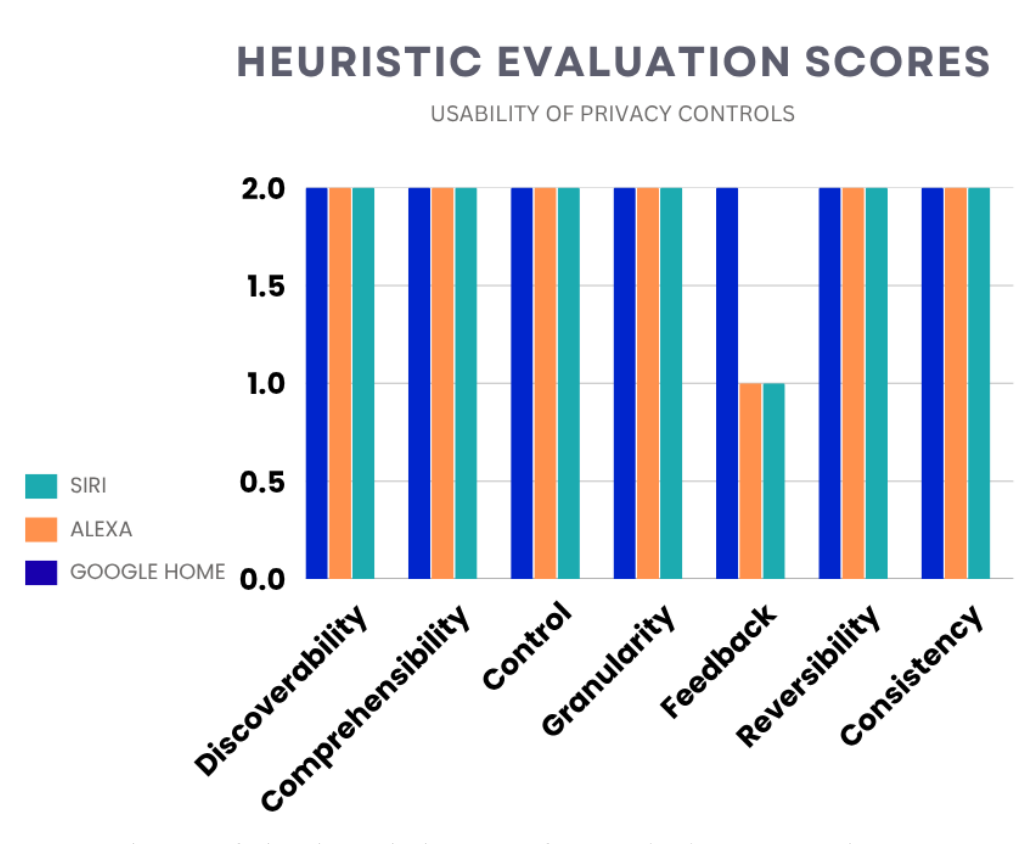

Fig. 1. Bar chart of the breakdown of Heuristic scores by smart device

TABLE III. UX TASK PERFORMANCE SUMMARY (EASE OF USE, NUMBER OF STEPS, TIME TAKEN)

| Task | Ease Of Use | | | Number of Steps | | | Time Taken | | |
|---|---|---|---|---|---|---|---|---|---|
| | *Google Home* | *Alexa* | *Siri* | *Google Home* | *Alexa* | *Siri* | *Google Home* | *Alexa* | *Siri* |
| Disable Voice History | 3.5 | 4.3 | 2.5 | 5 | 5 | 5 | 95 | 103 | 140 |
| Find Data Sharing Info | 4.3 | 4.5 | 4.3 | 3 | 3 | 3 | 80 | 41 | 100 |
| Delete Commands | 4.3 | 4.8 | 4.7 | 5 | 4 | 4 | 95 | 38 | 100 |
| Verify Change | 4.3 | 4.7 | 4.5 | 5 | 5 | 3 | 50 | 29 | 60 |

UX Task Performance Summary for Youth Participants (Average Ease 1–5, Average Number of Steps Rounded to the Nearest Integer, Average Time Taken in Seconds)

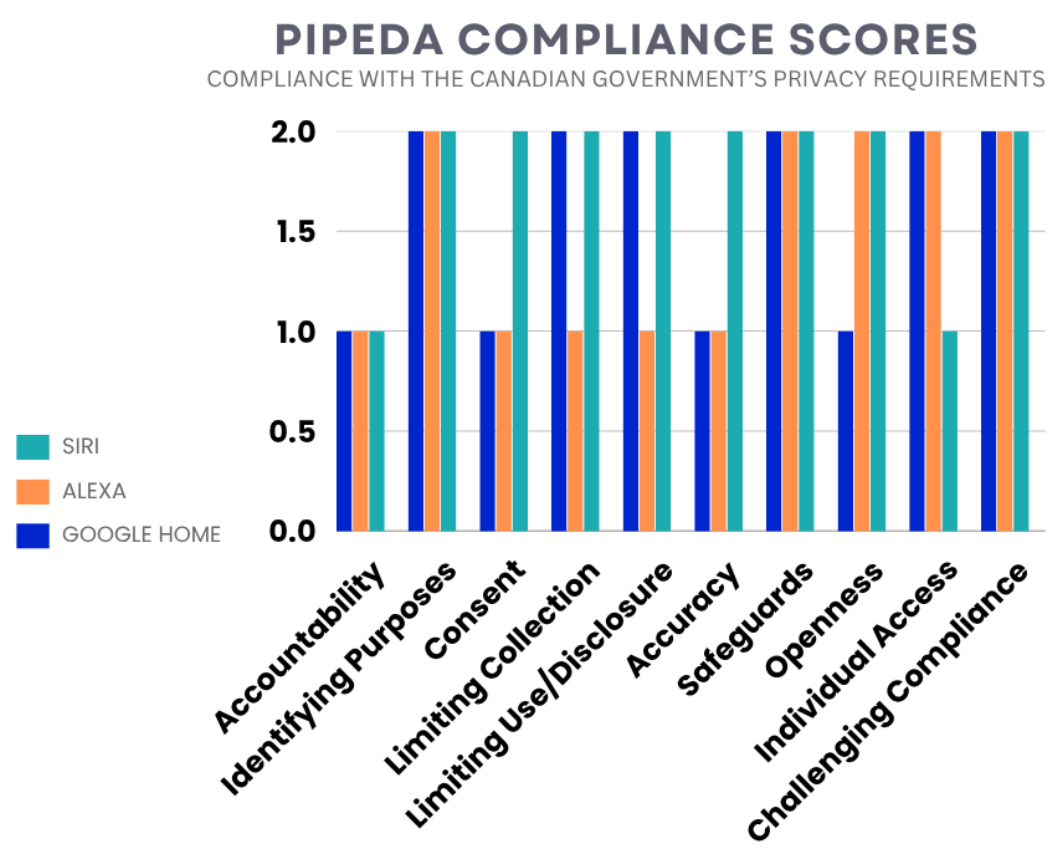

Fig. 2. Bar chart of the breakdown of PIPEDA scores by smart device

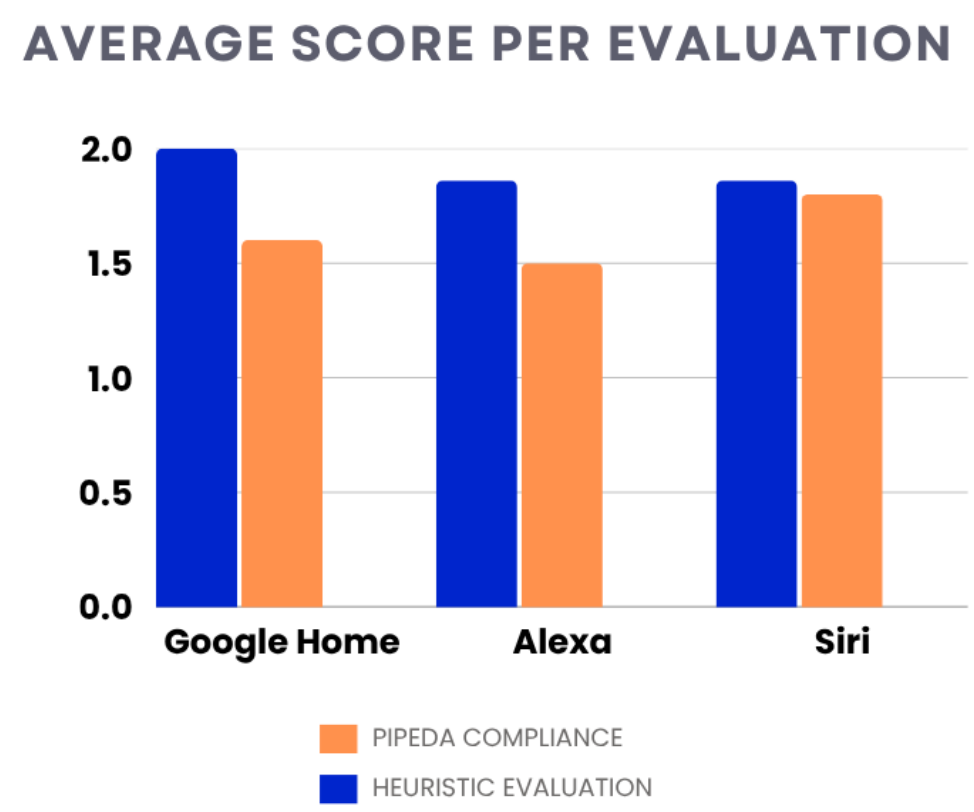

Fig. 3. Comparative Bar Chart (Heuristic vs. PIPEDA Compliance)

Similar to Google, Siri's setting to disable voice recording history was located outside of Siri's dedicated settings, which made it difficult for participants to locate. Lastly, Amazon's Alexa led in perceived ease of use (distinct from overall usability) by placing its settings in a more intuitive location compared to the other two devices. Nonetheless, during UX testing, it was discovered that disabling voice recording history on Alexa could take up to 36 hours to take effect. This finding highlighted that Amazon retains voice recordings and their transcripts indefinitely unless users manually configure an auto-delete setting or delete the data themselves. Fig. 3 illustrates the comparative performance across heuristic and compliance dimensions, revealing a trade-off between usability and regulatory alignment.

## V. DISCUSSION

### A. *Privacy Self-Efficacy and Meaningful Control*

Data privacy in smart devices has become a growing public concern, particularly with the rapid integration of AI technologies in domestic environments [48]. Although these devices offer several advantages, such as convenience, automation, and accessibility [48], [49], [50], young people still face significant challenges in understanding how their personal data is collected, stored, and shared [51], [52]. Many youths rely on smart devices such as Siri, Google Home, and Alexa for everyday communication, entertainment, or study without being fully aware of the extent of data processing happening in the background. This creates a gap between convenience and control, where their reliance on technology outweighs their perceived ability to protect their privacy [48].

In terms of privacy self-efficacy, this means that while youth may feel confident in their ability to understand, manage, and protect their personal information [53], their actual control over data remains limited by the hidden and complex nature of privacy settings that exist in these AI-driven systems. Psychologically, this gap can lead to feelings of helplessness or "privacy fatigue", where young people believe that protecting their data is impossible, no matter what they do [48], [54]. Socially, privacy choices are influenced by peer norms and social media behaviors, where sharing and visibility are often encouraged over caution [51]. As AI systems evolve, the amount of personal and behavioral data extracted from users expands, raising concerns over whether youths can exercise meaningful agency in data management [48]. Addressing this challenge requires educational interventions, simplified controls, and regulatory measures that enhance youth confidence and capability in managing privacy [49], [53].

These dynamics were visible in our UX observations. For example, one participant attempting to disable voice history on Google Home reported "clicked into the wrong sub menu" twice and noted that it was "not entirely clear where it would be," while another described Siri's controls as being "in multiple places," making it "a little hard to navigate." On Alexa, participants expressed confusion that privacy changes "take 36 hours to take into effect," which undermined their confidence that any single action would truly stop retention. Such comments illustrate how interface complexity and delayed effects can contribute to feelings of confusion and privacy fatigue, even among motivated youth who are actively trying to manage their settings.

### B. Design Strengths and Weaknesses

The comparative audit revealed distinct design strengths and weaknesses across the three platforms. Siri demonstrated clear strengths in the use of plain and understandable language within its privacy interface, which made it easier for participants to comprehend what each setting controlled. Alexa, on the other hand, stood out for its task clarity, as participants found the layout of its setting more intuitive and quicker to navigate compared to other devices. However, several weaknesses were also observed across all three systems. Both Google Home and Siri had privacy settings that were coupled with broader account-level controls, which made it harder for users to identify where specific privacy options were located. Another weakness, especially evident in Alexa, was the opacity surrounding data retention [42], [55]. Participants discovered that even after disabling voice recording history, Amazon continued to retain data for an extended period unless users manually intervened. These patterns show that while some design elements support user understanding, others still obscure full transparency and limit meaningful control. This underscores the importance of PbD for data protection and transparency [20].

These observations point to several concrete design guidelines. First, core privacy controls, such as voice-history management, deletion, and data-sharing preferences, should not be scattered across multiple, conceptually unrelated menus; instead, they should be co-located in a dedicated privacy hub for each assistant. Second, plain-language descriptions of settings, as seen in Siri, should be paired with explicit indicators of retention periods and deletion effects (e.g., "recordings are kept for 18 months" or "transcripts are deleted within 24 hours"). Third, where account-level and device-level controls are intertwined, interfaces should clearly label the scope of each change ("applies to this device only" versus "applies to all devices linked to this account") to reduce ambiguity about consequences for the broader ecosystem.

### C. Trade-off Between Usability and Compliance

Results revealed a noticeable trade-off between usability and regulatory compliance. Google Home achieved its highest usability score owing to its responsive feedback popups, which improved the experience when changing settings. However, it did not perform as strongly in the PIPEDA compliance checklist, indicating weaker safeguards and consent mechanisms. Siri, in contrast, excelled in compliance by providing stronger consent transparency and greater accuracy in managing user data, but received lower usability ratings due to more difficult navigation and the separation of certain controls from Siri's main settings. Alexa performed best at ease of use but had lower compliance scores, mainly due to unclear data retention practices. This inverse relationship suggests that devices emphasizing convenience may underperform on regulatory alignment, whereas systems designed for strict compliance often impose greater navigational complexity. Achieving both remains a central challenge for developers implementing PbD.

Although our findings highlight a tension between usability and compliance with current products, this trade-off is not necessarily inherent. Other domains offer examples where strong privacy protections coexist with relatively simple interfaces, such as mobile operating-system permission dashboards and browser privacy controls that surface high-level summaries with drill-down options for advanced users. In principle, smart-device ecosystems could adopt similar patterns, for instance, combining clear, in-context consent prompts and retention summaries with streamlined toggles and defaults that favor minimization. The results of this audit, therefore, suggest that the observed trade-off reflects prevailing design and business choices rather than an unavoidable constraint, and that there is room for patterns that jointly optimize usability and regulatory alignment.

### D. Alignment With Existing Literature

These findings align with existing research highlighting persistent PbD gaps in smart voice-enabled devices. Studies have shown that while users may assume these systems are private by default, the actual configurations often favor data collection and long-term retention. The audit results reflect this same pattern in most of the devices that require multiple steps to locate privacy settings, and some features, such as disabling voice history, were buried within broader account options. This reinforces the idea that the current design of smart assistants prioritizes functionality and personalization over user autonomy, leaving youth with limited real control, even when options appear available.

### E. Design Responsibility and Transparency in Data Flows

The findings highlight the responsibility of manufacturers to make privacy more accessible and understandable from the very beginning of device use. Instead of requiring users to search for settings after setups, privacy guidance should be integrated directly into onboarding flows. This could include contextual explanations, plain-language summaries, and confirmation prompts. The higher feedback score achieved by Google Home illustrates how immediate confirmation enhances trust; embedding similar features in setup sequences could strengthen privacy self-efficacy.

As illustrated in Fig. 4, all three devices follow a comparable data-processing pathway, from local capture to encrypted transmission to cloud storage and retention to user-initiated deletion, yet differ in the transparency offered at each stage. Google Home and Alexa both store audio and transcripts in the cloud and use the resulting profiles for personalization and advertising, yet provide only limited, policy-level explanations of retention periods. Siri processes more content on-device and associates cloud-stored transcripts with a rotating identifier that is later disassociated but still relies on a combination of in-app text and web documentation to explain these mechanisms. In all three cases, LEDs or visual waveforms indicate when recording is active, but there are no equivalent indicators for when data are being retained, shared, or purged. Mapping the devices onto this flow makes clear that transparency is concentrated at the point of capture, whereas retention and sharing stages remain largely invisible.

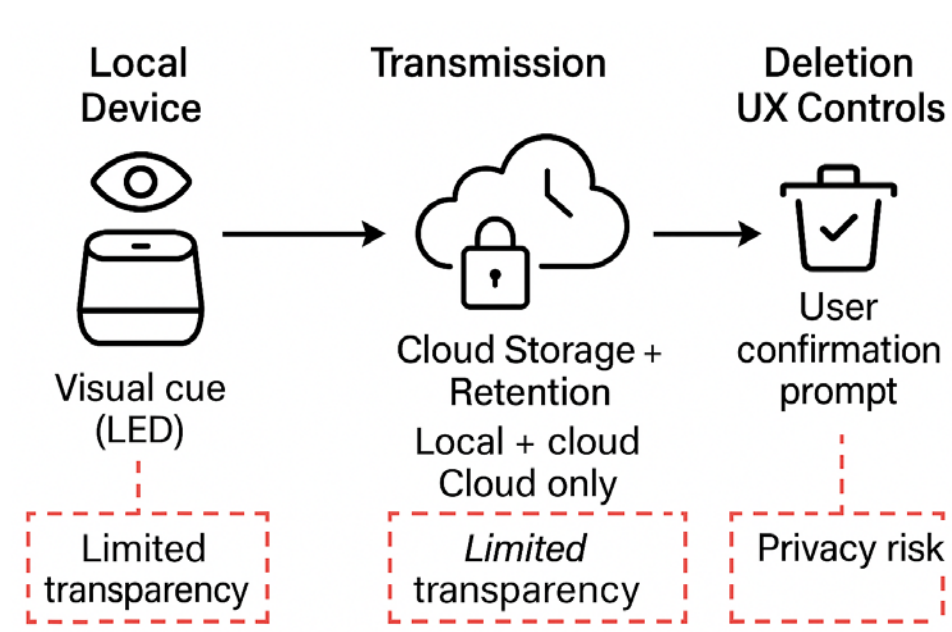

Fig. 4. Data Flow Diagram (collection–cloud–retention–deletion)

### F. Limitations and future work

This study has several limitations that also suggest directions for future research. The participant sample was intentionally small, consistent with formative usability testing norms, but this constrains generalizability. Additionally, the performance and interface layout of these devices may vary depending on firmware versions and regional updates, potentially influencing the outcomes observed in the heuristic and compliance evaluations. Furthermore, the PIPEDA compliance scores are based on expert judgment using a structured but bespoke checklist. Although a consensus-based procedure was used to reconcile differences, no formal inter-rater reliability statistics were calculated, and the resulting scores should therefore be interpreted as indicative rather than definitive measures of legal compliance.

Future studies should expand the scope of this audit to include a broader range of smart devices and platforms, as well as diverse user demographics and longitudinal engagement. Integrating surveys or telemetry-based analyses could help quantify how privacy behaviors evolve over time. Moreover, developing standardized audit frameworks for AI systems, combining heuristic usability metrics with compliance scoring, would enhance consistency and accountability across manufacturers.

Despite these constraints, the findings provide practical insight into how usability, compliance, and youth privacy self-efficacy intersect in AI-enabled smart ecosystems. Continued empirical evaluation is essential to ensure that PbD principles evolve from conceptual ideals into verifiable standards embedded directly within user experiences.

## VI. Conclusion

This paper presented a comparative PbD audit of three major AI-enabled smart devices, Google Home, Amazon Alexa, and Apple Siri, examining their usability, transparency, and compliance with Canada's PIPEDA framework. The primary contribution is an integrated PbD audit methodology that combines heuristic evaluation of privacy UX, a PIPEDA-based compliance checklist, and youth-centered task analysis, providing a holistic view of how privacy protections manifest in both interface design and lived user experience. The findings show that while all three devices integrate elements of PbD, implementation remains inconsistent and heavily reliant on users to adjust default settings. Among the platforms, Google Home achieved the highest usability score, supported by relatively clear feedback and intuitive navigation, yet overlapping account- and device-level controls introduced confusion and potential overreach. Alexa offered the most straightforward task navigation for common privacy actions, but this strength was undermined by opaque retention practices and delayed enforcement of deletion requests. Siri demonstrated the strongest overall alignment with fair information-handling principles, particularly in consent transparency and data accuracy, but its privacy settings were fragmented across menus, reducing practical discoverability and ease of use for youth. Across devices, three recurring challenges were identified: low discoverability of essential privacy controls, the dominance of opt-out consent models, and weak user feedback confirming when privacy choices take effect. These findings reveal an industry still balancing functionality, usability, and regulatory compliance rather than integrating them seamlessly. To advance user trust, particularly among youth, privacy must be redefined not as an optional configuration but as a core usability feature: transparent, verifiable, and effortless by design.

## Acknowledgment

This project has been funded by the Office of the Privacy Commissioner of Canada (OPC); the views expressed herein are those of the authors and do not necessarily reflect those of the OPC.